\documentclass[journal, 10pt, twocolumn]{IEEEtran}
\usepackage{epsfig}
\usepackage{amsmath,amssymb,amstext,amsthm}
\usepackage{epstopdf}
\usepackage{graphicx} 
\usepackage{ifthen}
\usepackage{psfrag}

\usepackage{subcaption}

\begin{document}
\title{Simultaneous Control Information and Power Transmission for Reconfigurable Intelligent Surfaces}
\author{Steven Kisseleff,~\IEEEmembership{Member,~IEEE}, Konstantinos Ntontin,~\IEEEmembership{Member,~IEEE}, Wallace A. Martins,~\IEEEmembership{Senior Member,~IEEE}, Symeon Chatzinotas,~\IEEEmembership{Senior Member,~IEEE} and Björn Ottersten,~\IEEEmembership{Fellow,~IEEE}\\
Interdisciplinary Centre for Security, Reliability and Trust (SnT), University of Luxembourg, Luxembourg.\\ E-mails: \{steven.kisseleff, konstantinos.ntontin, wallace.alvesmartins, symeon.chatzinotas, bjorn.ottersten\}@uni.lu.
\thanks{This work was supported by the Luxembourg National Research Fund (FNR) in the framework of the FNR CORE project RISOTTI C20/IS/14773976.}
}
\maketitle
\begin{abstract}
Reconfigurable intelligent surfaces (RISs) are planar structures with attached electronic circuitry that enable a partially programmable communication environment. RIS operation can be regarded as nearly passive since it acts by simply reflecting the impinging traveling waves towards desired directions, thus requiring energy only for the reconfiguration of its reflective elements (REs). This paper tackles the problem of wirelessly powering RIS circuitry via control signaling. Simultaneous wireless information and power transfer (SWIPT) is considered by taking into account two basic principles: that signal quality of the control signals is sufficient for information detection, and that there is enough harvested energy for the reconfiguration. Some of the most common SWIPT receivers (time sharing, power splitting, dynamic power splitting, and antenna selection) are studied and the corresponding proposed optimization problems implementing the aforementioned principles are formulated and solved in closed form. Numerical results show the effectiveness of the proposed methods in the presence of received power fluctuations.

\end{abstract}

\section{Introduction} 
\label{Introduction}
Reconfigurable intelligent surfaces (RISs) are an
emerging and promising technology, which can be used to increase the spectral efficiency and coverage of wireless networks. The reconfigurable properties of RIS are enabled via integrated electronic circuits, which can be programmed to reflect an impinging electromagnetic wave in a controlled manner \cite{liaskos2018new}. 
The benefits of RIS for wireless communication have been analyzed for various scenarios and applications, including 6G \cite{basar2020reconfigurable}, smart cities \cite{kisseleff2020reconfigurable}, challenging environments \cite{kisseleff2021reconfigurable}, satellite communications \cite{tekb2021reconfigurable}, etc. In these applications, notable improvements in data rate and energy efficiency can be expected. 

One of the advantages of RIS over traditional relaying is its low power consumption, since no active signal retransmission is needed \cite{di2020reconfigurable}. In fact, the consumed power is not linked to the power of the impinging signal, but rather to the internal calculations and updates of the impedances. Furthermore, potential applications of RIS may require autonomous and perpetual operation of RIS devices \cite{kisseleff2020reconfigurable}. For this, RIS needs to receive both the control information and the energy to be able to update the respective impedances. In this context, the use of simultaneous wireless information and power transfer (SWIPT) technology seems promising. 

SWIPT is a method of utilizing the same radio frequency (RF) signal for information and energy transfer. There are different configurations of SWIPT, which have been thoroughly investigated in the past~\cite{perera2017simultaneous}. In the context of RIS, typically only RIS-assisted SWIPT has been considered, where both the information and the energy are intended for the users of a communication system~\cite{pan2020intelligent,wu2020joint}. Recently, the concept of 'self-sustainable' RIS has been proposed, where part of the energy of the impinging signal is utilized for the powering of RIS~\cite{zou2020wireless,lyu2020optimized,pan2021selfsustainable}. However, in these works, the transmitted information is not intended for RIS, but for the users.

In this work, we propose to employ SWIPT technology in order to simultaneously provide control information and energy to the RIS device using the same signal. Thus, the maintenance of RIS would be decoupled from its main operation, which is very beneficial from the practical perspective. In particular, the optimization of the reflective elements (REs) would not need to account for any internal operation of RIS and its demands. Through this, some of the optimization constraints can be relaxed, potentially leading to  performance improvement and complexity reduction. Furthermore, the resulting technical solution may become less vulnerable to the channel estimation errors and thus to the performance degradation. Accordingly, the powering of RIS using the control signal transmitted separately from the network data is very beneficial for the design of RIS-assisted networks.

The contributions of this work are summarized as follows:
\begin{itemize}
    \item We consider four most common SWIPT methods and provide a generic architecture for simultaneous control signal and power transfer for RIS devices;
    \item For each selected SWIPT method we formulate an optimization problem with the objective to maximize the number of updated REs;
    \item We provide closed-form solutions for all four optimization problems and evaluate their performance;
    \item For a practical scenario, we investigate the impact of receive power fluctuations and provide a simple method to compensate the performance degradation.
\end{itemize}

This paper is organized as follows. In Section~\ref{Model}, the employed signal propagation model is described. Based on this model, four optimization problems corresponding to different SWIPT methods are formulated and analytically solved in Section~\ref{Solution}. Section~\ref{Results} shows the numerical results and Section~\ref{Conclusion} concludes the paper. 
\section{System model}
\label{Model}
\subsection{Signal model}
We consider a downlink of a multi-user multiple-input single-output (MU-MISO) system with a base station (BS) using $N$ antennas and assisted by a single RIS with $L_{\max}$ REs. In this work, we focus on the receiving module at the RIS, such that the distribution of users in the network, their demands and channels are not important for the considered scenario. For the RIS, we assume an integrated architecture \cite{abadal2020programmable}, i.e. each RE is connected to a controller, which is responsible for updating its impedance. In order to provide the control signal and power to the RIS, the BS uses a separate communication channel, which does not interfere with the downlink or uplink transmissions in the network. 

We describe the communication link between the BS and RIS as a single-user multiple-input multiple-output (MIMO) channel with a complex-valued channel matrix $\textbf{H}\in\mathbb{C}^{L_{\max}\times N}$, since each reflector can be viewed as an antenna with signal reception capabilities according to \cite{zou2020wireless}. Specifically, the signal absorbed by each RE can be used for information detection (ID) and energy harvesting (EH). Each link channel between a BS antenna and an RE also includes the loss associated with the signal reflection from this RE. The channel matrix $\textbf{H}$ is assumed to be known to the BS and RIS either from the previous estimations or based on their geographical location.

We denote the transmit signal vector as $\textbf{x}(t)=\textbf{a}s(t)$ with $\textbf{a}\in\mathbb{C}^{N\times 1}$, $\|\textbf{a}\|_2^2=1$ and control signal $s(t)$. 
The control signal $s(t)$ represents a data packet, which is composed of a sequence of $L_{\max}$ sub-packets of length $T$. Each sub-packet carries the update information for the respective RE. The transmit power is given by $P_{\rm t}=\mathcal{E}\{\left|s(t)\right|^2\}$, where $\mathcal{E}\{\cdot\}$ is the expectation operator.
The receive signal vector is represented by $\textbf{y}(t)\in\mathbb{C}^{L_{\max}\times 1}$. Moreover, the received signal at each RE is impaired by additive white Gaussian noise (AWGN) $\textbf{w}(t)$ with variance $\sigma^2$, i.e. $\mathcal{E}\{\textbf{w}(t)\textbf{w}(t)^{\rm H}\}=\sigma^2\cdot\textbf{I}$ with identity matrix $\textbf{I}$. The signal propagation can be thus described via
\begin{equation}
    \textbf{y}(t)=\textbf{H}\textbf{x}(t)=\textbf{H}\textbf{a}s(t)+\textbf{w}(t).
\end{equation}

\begin{figure}
    \centering
    \includegraphics[width=0.48\textwidth]{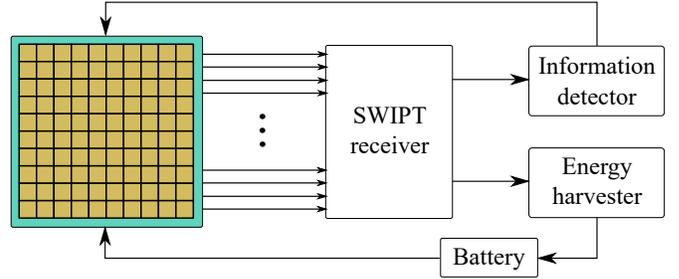}
    \caption{SWIPT-enabled receiver of control signals for RIS.}
    \label{fig:SWIPT}
    \vspace*{-2mm}
\end{figure}

We assume that all REs are connected to a single SWIPT receiver. Note that it seems impractical to connect each RE to an individual SWIPT receiver due to a large number of REs, such that the resulting implementation would become very complex and bulky. In order to not implicitly inject power into the system through signal amplification, we apply an equal gain combining (EGC) strategy \cite{brennan1959linear} at the RIS, which leads to $\exp{\left(\rm j\mathbf{\Theta}\right)}^{\rm H}\textbf{y}(t)$, where ${\rm j}=\sqrt{-1}$ is an imaginary unit and $\mathbf{\Theta}$ contains the phases of signal vector $\textbf{H}\textbf{a}$. 
The received power of the useful signal can be expressed as\footnote{One of the SWIPT methods, namely antenna selection method, requires a splitting of the signals in two parts before the summation. This can be achieved via $\exp{\left(\rm j\mathbf{\Theta}\right)}^{\rm H}\textbf{Q}\textbf{y}(t)$, where $\textbf{Q}$ is a  diagonal matrix, which contains `1`s on the main diagonal corresponding to the indices of REs selected for the intended SWIPT operation, i.e. EH or ID, and `0`s otherwise.} 
\begin{equation}
    \label{eq:pr}
    P_{\rm r}=\mathcal{E}\left\{\left|\exp{\left(\rm j\mathbf{\Theta}\right)}^{\rm H}\textbf{H}\textbf{a}s(t)\right|^2\right\}=\left|\exp{\left(\rm j\mathbf{\Theta}\right)}^{\rm H}\textbf{H}\textbf{a}\right|^2P_{\rm t}
\end{equation}
The choice of $\textbf{a}$ maximizing both the signal quality and the harvested power is a maximum ratio transmission (MRT) filter \cite{lo1999maximum} given by $\textbf{a}=\frac{\textbf{H}^{\rm H}\exp{\left(\rm j\mathbf{\Theta}\right)}}{\|\textbf{H}^{\rm H}\exp{\left(\rm j\mathbf{\Theta}\right)}\|_2}$. Correspondingly, we can express $P_{\rm r}=\|\textbf{H}^{\rm H}\exp{\left(\rm j\mathbf{\Theta}\right)}\|_2^2P_{\rm t}$ by inserting $\textbf{a}$ in \eqref{eq:pr}. 
The receiver consists of a SWIPT receiver, which implements the selected SWIPT method, energy harvester connected to a battery and information detector, see Fig.~\ref{fig:SWIPT}.
The received data is decoded while the energy is stored in the battery and then used to update the REs according to the decoded data.
\subsection{RIS receiver architecture}
For the SWIPT splitter, we consider four methods, which seem to be very promising for this application:\footnote{We employ various SWIPT methods with different definitions of splitting/sharing factors. Hence, the notation may deviate from the conventional works, which focus on a single SWIPT method.}
\begin{itemize}
    \item time sharing (TS) with sharing factor $\alpha$,
    \item power splitting (PS) with splitting factor $\rho$,
    \item dynamic power splitting (DS) with splitting factor $\gamma(\ell)$,
    \item antenna selection (AS) with subset cardinality $\eta$.
\end{itemize}
In TS, some of the received sub-packets will be used for EH and the others for ID. We introduce a binary variable $\tau(\ell)$ in order to indicate this. Note that $\alpha$ depends on $\tau(\ell)$ according to $\alpha=\frac{1}{L_{\max}}\sum_{\ell=1}^{L_{\max}}\tau(\ell)$. 
In PS and DS, the received power is split using a specifically designed power splitter. In AS, the set of antennas is split into two subsets, such that the signals from each subset of antennas are (coherently) combined. 
For more information about SWIPT receivers, please refer to \cite{perera2017simultaneous}.

For DS, we assume that $\gamma(\ell)$ can be modified after the reception of the $\ell$th sub-packet, similar to \cite{liu2013wireless}. Note that DS can be viewed as a joint application of TS and traditional PS. Specifically, TS can be achieved, if $\gamma(\ell)$ is set to zero or one for the respective sub-packets. 
\subsection{Energy harvesting and information detection}
For the input powers $P_{\mathrm{harv}}(\ell)$ and $P_{\mathrm{info}}(\ell)$ of EH and ID circuits, respectively, in $\ell$th sub-packet, we consider
\begin{equation}
\begin{tabular}{lll}
    TS: & $P_{\mathrm{harv}}(\ell)=P_{\rm r}\tau(\ell)$, & $P_{\mathrm{info}}(\ell)=P_{\rm r}(1-\tau(\ell))$, \\
    PS: & $P_{\mathrm{harv}}(\ell)=P_{\rm r}\rho$, & $P_{\mathrm{info}}(\ell)=P_{\rm r}(1-\rho)$, \\
    DS: & $P_{\mathrm{harv}}(\ell)=P_{\rm r}\gamma(\ell)$, & $P_{\mathrm{info}}(\ell)=P_{\rm r}(1-\gamma(\ell))$,\\
    AS: & $P_{\mathrm{harv}}(\ell)=P_{\rm r}\eta^2$, & $P_{\mathrm{info}}(\ell)=P_{\rm r}(1-\eta)^2$.
\end{tabular}
\label{eq:energy_info}
\end{equation}
For EH, we assume a non-linear model according to \cite{boshkovska2015practical}, such that the corresponding harvested energy is given by
\begin{equation}
    E_r(\ell)=\frac{\hat{E}}{1-\phi}\left(\frac{1}{1+\operatorname{e}^{-q P_{\mathrm{harv}}(\ell)+qr}}-\phi\right),
\label{eq:E_r}
\end{equation}
with $\phi=\frac{1}{1+\exp(qr)}$, where $\hat{E}$, $q$ and $r$ are the specific parameters of the rectifying circuit.
We denote the maximum harvested energy, if all power $P_{\rm r}$ is dedicated to EH, as $E_{\max}$.
In this work, we assume that a certain minimum energy $E_0$ is required in order to be able to update the reflective properties of a single RE. This is a reasonable assumption, since each RE is connected to a steerable impedance, which needs to be manipulated in order to change the reflective properties of the RE. Additional static power consumption during main operation of RIS may need to be compensated as well. However, this additional static power is typically very low, as mentioned earlier, and thus negligible.

The signal quality in the $\ell$th sub-packet can be expressed in terms of signal-to-noise ratio (SNR). For TS and AS, the SNR is given by $\mathrm{SNR}(\ell)=\frac{P_{\mathrm{info}}(\ell)}{\sigma^2}$ while for PS and DS we assume that the signal is further impaired by an additional AWGN with variance $\delta^2$, which is injected by the employed power splitter \cite{shi2014joint}. 
We denote the maximum possible signal quality, i.e. if all power $P_{\rm r}$ is used for ID, as $\mathrm{SNR}_{\max}=\frac{P_{\rm r}}{\sigma^2}$. The minimum required SNR necessary for a correct detection of the data is denoted as $\mathrm{SNR}_{0}$. We assume that an update is only possible if the actual signal quality is above this value.
\vspace*{1mm}
\section{Control information and power reception}
\label{Solution}
In this section, we discuss multiple optimization problems corresponding to the four described SWIPT methods. Since the primary goal of RIS is the assistance of the wireless network in signal propagation and its efficiency is directly related to the number of properly optimized REs, the number of updated REs should be as high as possible. Correspondingly, in this work, we select the maximization of the number of updated REs as the objective in all considered optimization problems. Furthermore, we select $\alpha$, $\rho$, $\gamma(\ell)$ and $\eta$ as the sole optimization parameters in the respective problems.\\ 
Note that RIS can decide which elements should be updated, since the assignment of the sub-packets to the REs is known. Without any loss of generality, we assume that the first $L$ elements/sub-packets need to be updated. Moreover, we focus on the reception of a single control sequence, i.e., only the energy harvested during this sequence is considered. In practice, excess energy can be stored in the battery and used to update more REs in future update cycles. Correspondingly, the design may slightly vary depending on the stored energy. 
\subsection{Time sharing based SWIPT}
For TS, the sharing factor $\alpha$ determines how many sub-packets will be used to update the REs and how many sub-packets for EH. Note that with $\tau(\ell)=1$, all power of the sub-packet will be used for EH, i.e. $E_r(\ell)=E_{\max},\: \ell>L$. Hence, the total harvested energy is given by $\sum_{\ell=L+1}^{L_{\max}}E_r(\ell)=E_{\max}(L_{\max}-L)$. Furthermore, with $\tau(\ell)=0$, all power will be used for ID, i.e. $\mathrm{SNR}(\ell)=\mathrm{SNR}_{\max},\: \ell\leq L$. The optimization problem can be formulated as follows:
\begin{eqnarray}
\underset{L}{\rm maximize}&L,\notag\\
\mbox{subject to:} &1)& E_{\max}(L_{\max}-L)\geq LE_0,\\
&2)& \mathrm{SNR}_{\max}\geq \mathrm{SNR}_{0}.\notag
\label{eq:problem1}
\end{eqnarray}
The first constraint indicates that the harvested energy should be enough to update $L$ REs. The second constraint ensures that the signal quality is sufficient. Since $\mathrm{SNR}_{\max}$ is independent of $L$, the second constraint represents a feasibility check, i.e. no updates can be made, if $\mathrm{SNR}_{\max}< \mathrm{SNR}_{0}$ holds.

Since the maximization of $L$ is only restricted by constraint 1), in the optimum, this constraint is active. By reformulating it, we obtain an upperbound on $L$ given by
$L=\left\lfloor L_{\max}\frac{E_{\max}}{E_{\max}+E_0}\right\rfloor$.

\subsection{Power splitting based SWIPT}
The splitting factor $\rho$ determines the amount of power to be guided into the energy harvester. Correspondingly, $E_r(\ell)$ depends on $\rho$ according to \eqref{eq:E_r} and \eqref{eq:energy_info}. This power is extracted from the received signal power of all sub-packets, such that we obtain $E_r(\ell)=E_r(1),\:\forall \ell$. On the other hand, the signal quality after the splitting reduces due to the additional noise signal \cite{shi2014joint}. The resulting signal quality is therefore given by 
\begin{equation}
    \mathrm{SNR}(\ell)=\frac{P_{\mathrm{info}}(\ell)}{\sigma^2(1-\rho)+\delta^2},
\end{equation}
where $P_{\mathrm{info}}(\ell)$ depends on $\rho$ as stated earlier. Hence, we can formulate the following optimization problem:
\begin{eqnarray}
\label{eq:problem2}
\underset{0\leq\rho\leq1}{\rm maximize}&L,\notag\\
\mbox{subject to:} &1)& E_r(1)L_{\max}\geq LE_0,\\
&2)& \mathrm{SNR}(\ell)\geq \mathrm{SNR}_{0},\:\forall \ell\notag.
\end{eqnarray}
We obtain an expression with respect to $\rho$ by reformulating the second  constraint:
\begin{equation}
    \rho\leq\frac{P_{\rm r}-\mathrm{SNR}_{0}(\sigma^2+\delta^2)}{P_{\rm r}-\mathrm{SNR}_{0}\sigma^2}=\frac{1-(\frac{\mathrm{SNR}_{0}}{\mathrm{SNR}_{\max}}+\frac{\mathrm{SNR}_{0}}{\mathrm{SNR}_{\rm split}})}{1-\frac{\mathrm{SNR}_{0}}{\mathrm{SNR}_{\max}}},
    \label{eq:rho}
\end{equation}
where $\mathrm{SNR}_{\rm split}=\frac{P_{\rm r}}{\delta^2}$ denotes the signal quality experienced by the power splitter in absence of the original AWGN with variance $\sigma^2$. In the optimum, this expression becomes an equality in order to harvest as much energy as possible.

Then, we can obtain the optimal values for $P_{\mathrm{harv}}(1)$ and $E_r(1)$ by inserting \eqref{eq:rho} into \eqref{eq:energy_info} and then \eqref{eq:E_r}. Then, using the first constraint we obtain the upperbound 
    $L=\left\lfloor\frac{E_r(1)L_{\max}}{E_0}\right\rfloor$.
\vspace*{1mm}

\subsection{Dynamic power splitting based SWIPT}
For the dynamic power splitting, we can formulate the optimization problem by slightly modifying \eqref{eq:problem2}:
\vspace*{-1mm}
\begin{eqnarray}
\label{eq:problem3}
\underset{0\leq\gamma(\ell)\leq1,\:\forall \ell}{\rm maximize}&L,\notag\\
\mbox{subject to:} &1)& \sum_{\ell=1}^{L_{\max}}E_r(\ell)\geq LE_0,\\
&2)& \mathrm{SNR}(\ell)\geq \mathrm{SNR}_{0},\: 1\leq\ell\leq L\notag.
\end{eqnarray}
The second constraint is only needed for the first $L$ sub-packets, since the REs corresponding to all other sub-packets cannot be updated due to lack of energy. On the other hand, for the first $L$ sub-packets, this constraint should be fulfilled with equality in order to harvest as much energy as possible. Hence, we obtain for $\ell\leq L$ the same result as in \eqref{eq:rho}, i.e. $\gamma(\ell)=\rho$. The values of $P_{\mathrm{harv}}(\ell)$ and $E_r(\ell)$ can be determined accordingly. Since $E_r(\ell)$ remains constant for sub-packets $\ell\leq L$, we can express $\sum_{\ell=1}^{L}E_r(\ell)=LE_r(1)$. For sub-packets $\ell> L$, the information detection is not needed, such that we can set the respective power splitting factor $\gamma(\ell)=1,\:\ell> L$. Correspondingly,  $E_r(\ell)=E_{\max},\:\ell> L$ holds. Hence, the total harvested energy can be expressed as
\vspace*{-1mm}
\begin{equation}
    \sum_{\ell=1}^{L_{\max}}E_r(\ell)=LE_r(1)+(L_{\max}-L)E_{\max}.
    \vspace*{-1mm}
\end{equation}
By inserting this result in the first constraint of \eqref{eq:problem3}, we obtain
\vspace*{-1mm}
\begin{equation}
    L=\left\lfloor L_{\max}\frac{E_{\max}}{E_{\max}+E_0-E_r(1)}\right\rfloor.
\end{equation}
\subsection{Antenna selection based SWIPT}
In this method, the antennas or REs are split into two subsets/clusters pertaining either to ID or EH. Usually, only the subset cardinality is important, since the received powers are assumed to be independent of the antenna indices in the long run. For the co-located REs, the increased correlation of the channels and impedances (cf. \cite{vanchien2021outage}) leads to a similar result even for an instantaneous observation, since the received power is similar with all REs. Here, we assume that the channels between BS antennas and the REs are fully correlated, such that $\textbf{a}\approx \frac{\exp\left(\rm j\mathbf{\Phi}\right)}{\sqrt{L_{\max}}}$ holds with a phase vector $\mathbf{\Phi}$. 

Part of the received power $P_{\rm r}$ coming from $\eta L_{\max}$ REs will be harvested and part of $P_{\rm r}$ from $L_{\max}(1-\eta)$ REs will be used for ID. Note that $\eta L_{\max}$ is an integer value corresponding to a number of REs. The resulting optimization problem is formulated as follows:
\vspace*{-1mm}
\begin{eqnarray}
\label{eq:problem4}
\underset{0\leq\eta\leq1}{{\rm maximize}}&L,\notag\\
\mbox{subject to:} &1)& E_r(1)L_{\max}\geq LE_0,\\
&2)& \mathrm{SNR}(\ell)\geq \mathrm{SNR}_{0},\forall \ell\notag,\\
&3)& L_{\max}\eta\ \mathrm{integer}.\notag
\vspace*{-0.5mm}
\end{eqnarray}
At first, the signal quality is determined. For this, we consider that the useful signals from $L_{\max}(1-\eta)$ REs are coherently combined\footnote{Note that the useful signals from all antennas are given by the vector $\textbf{H}\textbf{a}\approx\textbf{H}\frac{\exp\left(\rm j\mathbf{\Phi}\right)}{\sqrt{L_{\max}}}\sim \exp{\left(\rm j\mathbf{\Theta}\right)}$. If only the $(1-\eta)$th portion of all REs is used, the magnitude of the coherent combination of signals is proportional to $(1-\eta)$. Hence, the received signal power scales with $(1-\eta)^2$.} leading to the receive power $P_{\rm r}(1-\eta)^2$ while the noise signals are not, which leads to the received noise power $\sigma^2(1-\eta)$. Hence, $\mathrm{SNR}(\ell)=\frac{P_{\rm r}(1-\eta)^2}{\sigma^2(1-\eta)}=\mathrm{SNR}_{\max}(1-\eta)$ holds. By equating this value with $\mathrm{SNR}_{0}$ and forcing $\eta L_{\max}$ to be an integer, we obtain
\begin{equation}
    \eta=\left\lfloor L_{\max}-\frac{\mathrm{SNR}_{0}}{\mathrm{SNR}_{\max}}L_{\max}\right\rfloor\frac{1}{L_{\max}}.
\end{equation}
With this result, we calculate $P_{\mathrm{harv}}=P_{\rm r}\eta^2$ and the harvested energy $E_r(1)$ from \eqref{eq:E_r}, which remains equal for all $L_{\max}$ sub-packets. Then, the maximum number of elements to be updated is determined by inserting $E_r(1)$ in $L=\left\lfloor\frac{E_r(1)L_{\max}}{E_0}\right\rfloor$.

\section{Numerical results}
\label{Results}
In this section, we show the results of numerical evaluations of the proposed methods. We select a transmission distance of 75 m between the BS and RIS. The RIS is assumed to have $L_{\max}=100$ elements with absorption loss of 3 dB. For the signal propagation, we assume a quasi-static frequency-flat channel with path loss coefficient of 3.5 and carrier frequency of 2.4 GHz. The non-linear EH parameters are given by $\hat{E}=2.8$ mJ, $q=1500$ and $r=0.0022$ \cite{guo2014theoretical}. AWGN variances are given by $\sigma^2=\delta^2=-100$ dBW. The requirements for the update of each RE are set to $\mathrm{SNR}_{0}=7$ dB and $E_0=1$ nJ. 

\begin{figure}
    \centering
    \includegraphics[width=0.48\textwidth]{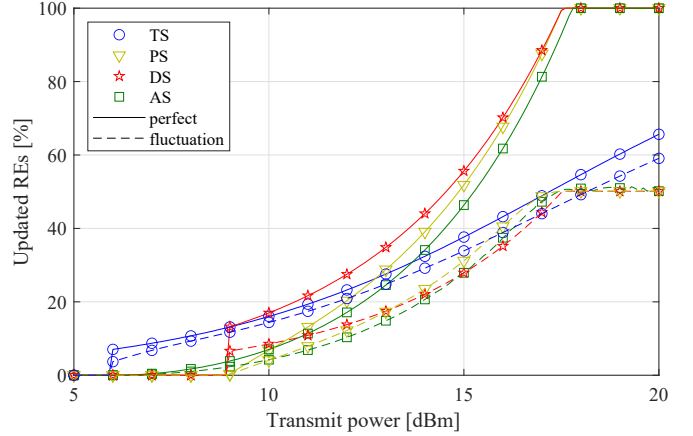}
    \caption{Relative number of REs vs. transmit power.}
    \label{fig:1}
    \vspace*{-2mm}
\end{figure}

We start by investigating the performance of SWIPT under perfect knowledge of receive power and under power fluctuations. For this, we assume that $P_{\rm r}$ follows the normal distribution with a standard deviation (std) of $0.25P_{\rm r}$ around the mean value of $P_{\rm r}$. Moreover, we average the results of $10^5$ transmissions of control sequences. For a correct assessment, we assume for all methods that the respective splitting/sharing parameters remain unchanged. As a side effect, the true number of updated REs with TS and DS cannot be larger than it was assumed with no power fluctuations, since the choice of $\tau(\ell)$ and $\gamma(\ell)$ dictates which sub-packets should be detected. On the other hand, with PS and AS it is possible to detect all sub-packets and decide later, how many REs will be updated based on collected energy. The results are depicted in Fig.~\ref{fig:1}.
We observe that TS outperforms other methods with low transmit power and in case of receive power fluctuations. With a perfect knowledge of $P_{\rm r}$, PS, DS and AS methods are much better at higher transmit powers reaching $100\%$ with 17.5 dBm. However, due to power fluctuations, we observe a performance degradation, which reduces the maximum number of updated REs to $50\%$. The reason is the symmetry of normal distribution. If the true value of $P_{\rm r}$ drops below the assumed one (in $50\%$ of cases), the signal quality using these methods (without a proper update) is likely to decrease below $\mathrm{SNR}_{0}$, such that no update can be done at all.

\begin{figure}
    \centering
    \includegraphics[width=0.48\textwidth]{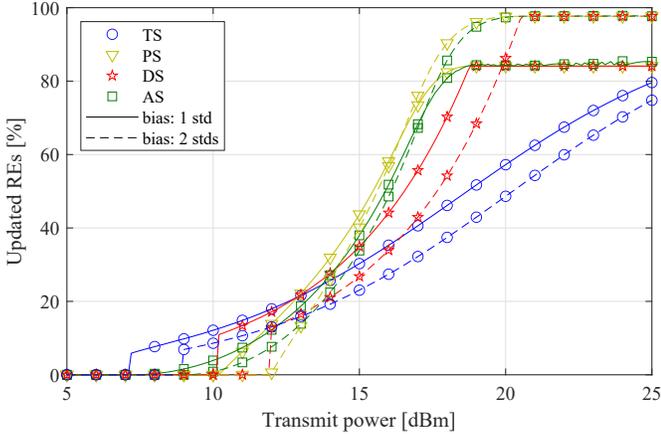}
    \caption{Impact of negative bias.}
    \label{fig:2}
    \vspace*{-2mm}
\end{figure}

The performance degradation can be reduced, if the received power distribution is not symmetric, but biased towards higher values of $P_{\rm r}$. This can be achieved by introducing a negative bias added to the assumed value of $P_{\rm r}$. We consider two possible values of the bias: (i) $0.25P_{\rm r}$ and (ii) $0.5P_{\rm r}$. These values correspond to 1 std and 2 stds of the normal distribution, respectively. The results are depicted in Fig.~\ref{fig:2}. 
We observe that TS and DS convergence is slower with 2 stds compared to 1 std bias. However, almost $100\%$ of updated REs can be reached with 2 stds bias using PS, DS and AS.

\begin{figure}
    \centering
    \includegraphics[width=0.48\textwidth]{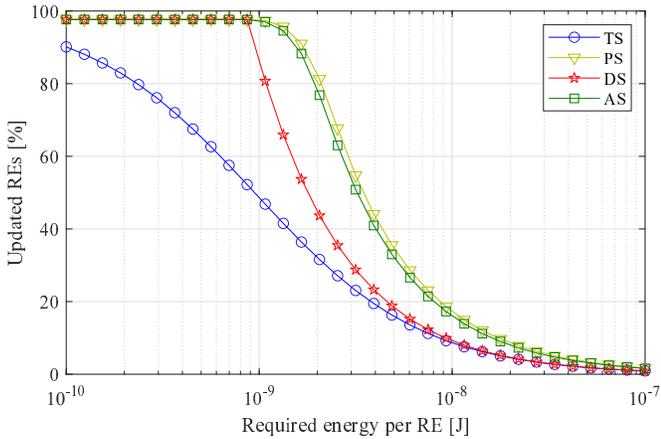}
    \caption{Impact of required energy $E_0$.}
    \label{fig:3}
    \vspace*{-2mm}
\end{figure}

Next, we investigate the dependency of the obtained solution on the required energy $E_0$ for updating a single RE. We assume a transmit power of 20 dBm and a negative bias of 2 stds. The results are shown in Fig. \ref{fig:3}. 
As expected, we observe a decrease of the number of updated REs with increasing $E_0$. Furthermore, we observe that PS and AS methods substantially outperform the other two methods -- a behavior already observed in Fig. \ref{fig:2}. 

\section{Conclusion}
\label{Conclusion}
In this work, we proposed a novel approach for updating the impedances of RIS using the control signals from the BS as information and energy carrier in order to account for the minimal signal quality and energy requirements. In this context, four typical SWIPT methods have been employed and the respective optimization problems have been formulated with the goal to maximize the number of REs to be updated using this control sequence. For each of the four problems, a closed-form solution has been derived. The numerical evaluations have shown that the PS and AS methods outperform the other two methods in case of perfectly known receive power and in case of partially compensated receive power fluctuations.
\bibliographystyle{IEEEtran}
\bibliography{Literature}

\begin{thebibliography}{10}
\providecommand{\url}[1]{#1}
\csname url@samestyle\endcsname
\providecommand{\newblock}{\relax}
\providecommand{\bibinfo}[2]{#2}
\providecommand{\BIBentrySTDinterwordspacing}{\spaceskip=0pt\relax}
\providecommand{\BIBentryALTinterwordstretchfactor}{4}
\providecommand{\BIBentryALTinterwordspacing}{\spaceskip=\fontdimen2\font plus
\BIBentryALTinterwordstretchfactor\fontdimen3\font minus
  \fontdimen4\font\relax}
\providecommand{\BIBforeignlanguage}[2]{{%
\expandafter\ifx\csname l@#1\endcsname\relax
\typeout{** WARNING: IEEEtran.bst: No hyphenation pattern has been}%
\typeout{** loaded for the language `#1'. Using the pattern for}%
\typeout{** the default language instead.}%
\else
\language=\csname l@#1\endcsname
\fi
#2}}
\providecommand{\BIBdecl}{\relax}
\BIBdecl

\bibitem{liaskos2018new}
C.~Liaskos, S.~Nie, A.~Tsioliaridou, A.~Pitsillides, S.~Ioannidis, and
  I.~Akyildiz, ``A new wireless communication paradigm through
  software-controlled metasurfaces,'' \emph{IEEE Communications Magazine},
  vol.~56, no.~9, pp. 162--169, 2018.

\bibitem{basar2020reconfigurable}
E.~Basar, ``{Reconfigurable intelligent surface-based index modulation: A new
  beyond MIMO paradigm for 6G},'' \emph{IEEE Transactions on Communications},
  vol.~68, no.~5, pp. 3187--3196, 2020.

\bibitem{kisseleff2020reconfigurable}
S.~Kisseleff, W.~A. Martins, H.~Al-Hraishawi, S.~Chatzinotas, and B.~Ottersten,
  ``Reconfigurable intelligent surfaces for smart cities: Research challenges
  and opportunities,'' \emph{IEEE Open Journal of the Communications Society},
  vol.~1, pp. 1781--1797, 2020.

\bibitem{kisseleff2021reconfigurable}
S.~Kisseleff, S.~Chatzinotas, and B.~Ottersten, ``Reconfigurable intelligent
  surfaces in challenging environments: Underwater, underground, industrial and
  disaster,'' \emph{arXiv:2011.12110}, 2021.

\bibitem{tekb2021reconfigurable}
K.~Tekbıyık, G.~K. Kurt, A.~R. Ekti, A.~Görçin, and H.~Yanikomeroglu,
  ``{Reconfigurable Intelligent Surfaces Empowered THz Communication in LEO
  Satellite Networks},'' \emph{arXiv:2007.04281}, 2021.

\bibitem{di2020reconfigurable}
M.~Di~Renzo \emph{et~al.}, ``Reconfigurable intelligent surfaces vs. relaying:
  Differences, similarities, and performance comparison,'' \emph{IEEE Open
  Journal of the Communications Society}, vol.~1, pp. 798--807, 2020.

\bibitem{perera2017simultaneous}
T.~D.~P. Perera, D.~N.~K. Jayakody, S.~K. Sharma, S.~Chatzinotas, and J.~Li,
  ``{Simultaneous wireless information and power transfer (SWIPT): Recent
  advances and future challenges},'' \emph{IEEE Communications Surveys \&
  Tutorials}, vol.~20, no.~1, pp. 264--302, 2017.

\bibitem{pan2020intelligent}
C.~Pan \emph{et~al.}, ``{Intelligent reflecting surface aided MIMO broadcasting
  for simultaneous wireless information and power transfer},'' \emph{IEEE J. on
  Sel. Areas in Commun.}, vol.~38, no.~8, pp. 1719--1734, 2020.

\bibitem{wu2020joint}
Q.~Wu and R.~Zhang, ``{Joint active and passive beamforming optimization for
  intelligent reflecting surface assisted SWIPT under QoS constraints},''
  \emph{IEEE Journal on Selected Areas in Communications}, vol.~38, no.~8, pp.
  1735--1748, 2020.

\bibitem{zou2020wireless}
Y.~Zou, S.~Gong, J.~Xu, W.~Cheng, D.~T. Hoang, and D.~Niyato, ``{Wireless
  Powered Intelligent Reflecting Surfaces for Enhancing Wireless
  Communications},'' \emph{IEEE Transactions on Vehicular Technology}, vol.~69,
  no.~10, pp. 12\,369--12\,373, 2020.

\bibitem{lyu2020optimized}
B.~Lyu, P.~Ramezani, D.~T. Hoang, S.~Gong, Z.~Yang, and A.~Jamalipour,
  ``{Optimized energy and information relaying in self-sustainable
  IRS-empowered WPCN},'' \emph{IEEE Transactions on Communications}, vol.~69,
  no.~1, pp. 619--633, 2020.

\bibitem{pan2021selfsustainable}
Y.~Pan, K.~Wang, C.~Pan, H.~Zhu, and J.~Wang, ``{Self-Sustainable
  Reconfigurable Intelligent Surface Aided Simultaneous Terahertz Information
  and Power Transfer (STIPT)},'' \emph{arXiv:2102.04053}, 2021.

\bibitem{abadal2020programmable}
S.~Abadal, T.-J. Cui, T.~Low, and J.~Georgiou, ``Programmable metamaterials for
  software-defined electromagnetic control: Circuits, systems, and
  architectures,'' \emph{IEEE Journal on Emerging and Selected Topics in
  Circuits and Systems}, vol.~10, no.~1, pp. 6--19, 2020.

\bibitem{brennan1959linear}
D.~G. Brennan, ``Linear diversity combining techniques,'' \emph{Proceedings of
  the IRE}, vol.~47, no.~6, pp. 1075--1102, 1959.

\bibitem{lo1999maximum}
T.~K. Lo, ``Maximum ratio transmission,'' in \emph{IEEE International
  Conference on Communications}, vol.~2.\hskip 1em plus 0.5em minus 0.4em\relax
  IEEE, 1999, pp. 1310--1314.

\bibitem{liu2013wireless}
L.~Liu, R.~Zhang, and K.-C. Chua, ``Wireless information and power transfer: A
  dynamic power splitting approach,'' \emph{IEEE Transactions on
  Communications}, vol.~61, no.~9, pp. 3990--4001, 2013.

\bibitem{boshkovska2015practical}
E.~Boshkovska \emph{et~al.}, ``{Practical non-linear energy harvesting model
  and resource allocation for SWIPT systems},'' \emph{IEEE Communications
  Letters}, vol.~19, no.~12, pp. 2082--2085, 2015.

\bibitem{shi2014joint}
Q.~Shi, L.~Liu, W.~Xu, and R.~Zhang, ``{Joint transmit beamforming and receive
  power splitting for MISO SWIPT systems},'' \emph{IEEE Transactions on
  Wireless Communications}, vol.~13, no.~6, pp. 3269--3280, 2014.

\bibitem{vanchien2021outage}
T.~V. Chien, A.~K. Papazafeiropoulos, L.~T. Tu, R.~Chopra, S.~Chatzinotas, and
  B.~Ottersten, ``{Outage probability analysis of IRS-assisted systems under
  spatially correlated channels},'' \emph{arXiv:2102.11408}, 2021.

\bibitem{guo2014theoretical}
J.~Guo, H.~Zhang, and X.~Zhu, ``{Theoretical analysis of RF-DC conversion
  efficiency for class-F rectifiers},'' \emph{IEEE transactions on microwave
  theory and techniques}, vol.~62, no.~4, pp. 977--985, 2014.

\end{thebibliography}
\end{document}